%% file: th.tex
\scrollmode

\documentclass[12pt,twoside,a4paper]{article}
\usepackage[dvips]{epsfig}
\newcommand{\jv}{{\tt JetViP}}
\newcommand{\al}{\alpha}

\newcommand{\g}{\gamma}
\newcommand{\G}{\Gamma}
\newcommand{\de}{\delta}

\newcommand{\e}{\epsilon}

\newcommand{\si}{\sigma}

\newcommand{\simgt}{\,\rlap{\lower 3.5 pt \hbox{$\mathchar \sim$}} \raise 1pt
 \hbox {$>$}\,}
\newcommand{\simlt}{\,\rlap{\lower 3.5 pt \hbox{$\mathchar \sim$}} \raise 1pt
 \hbox {$<$}\,}

\newcommand{\equ}[2]{\begin{equation} \label{#1} #2 \end{equation} }
\newtheorem{figgur}{Figure}[section]

\begin{document}
\thispagestyle{empty}
\title{\vskip-3cm{\baselineskip14pt
\centerline{\normalsize DESY 98-071 \hfill ISSN 0418--9833}
\centerline{\normalsize hep--ph/9806437 \hfill}
\centerline{\normalsize June 1998\hfill}
}
\vskip1.5cm \jv\ 1.1: \\ Calculating One- and Two-Jet
  Cross Sections \\ with Virtual Photons in NLO QCD \\
  \author{B.~P\"otter \\ II. Institut f\"ur Theoretische 
  Physik\thanks{Supported by Bundesministerium f\"ur Forschung und
  Technologie, Bonn, Germany, under Contract 05~7~HH~92P~(0),
  and by EEC Program {\it Human Capital and Mobility} through Network
  {\it Physics at High Energy Colliders} under Contract
  CHRX--CT93--0357 (DG12 COMA).}, Universit\"at Hamburg\\
  Luruper Chaussee 149, D-22761 Hamburg, Germany\\
  e-mail: poetter@mail.desy.de} }
\date{}
\maketitle

\begin{center}
{\bf Abstract}
\end{center}

\begin{quote}
\jv\ is a computer program for the calculation of inclusive single- and
dijet cross sections in $eP$- and $e\g$-scattering in NLO QCD. The
virtuality of the photon, radiated by the incoming electron, can be
chosen in a continuous range, reaching from photoproduction
into deep inelastic scattering. The various contributions to the full
jet cross section, including the resolved photon contributions,  
are implemented. The calculation is based on the phase-space-slicing
method.
\end{quote}

\newpage

\begin{center}
PROGRAM SUMMARY
\end{center}
{\em Titel of the program:} \jv\ version 1.1. \\
\\
{\em Computer-Enviroment:} Any machine running standard {\tt FORTRAN 77}.
\jv\ has been especially tested on HP-Clusters, IRIX-machines and
Linux PCs. \\  
\\
{\em Programming language used:} {\tt FORTRAN 77}. \\
\\
{\em High speed storage required:} Size of executable program is
approximately 9 MBytes. \\
\\
{\em Other programs used:} {\tt VEGAS} (multidimensional Monte Carlo
integration routine), {\tt PDFLIB, SaSgam, GRS} (parametrizations of
parton densities). 
\\
{\em Nature of physical problem:} In $eP$- and $e\g$-scattering
experiments, the hadronic final state can be analysed by jet cluster
algorithms, yielding inclusive single- and dijet cross sections. These
can be obtained in a continuous range of photon virtuality. The cross
sections allow the extraction of parameters, such as $\al_s$,
$\Lambda_{\overline{\mbox{MS}}}$ or parton densities (also of the virtual 
photon), if the respective jet cross sections are theoretically known. \\
\\
{\em Method of solution:} \jv\ is a computer program for the calculation 
of inclusive single- and
dijet cross sections in $eP$- and $e\g$-scattering in NLO QCD. The
virtuality of the photon, radiated by the incoming electron, can be
chosen in a continuous range, reaching from photoproduction
into deep inelastic scattering. The various contributions to the full
jet cross section, including the resolved photon contributions,  
are implemented. The calculation is based on the phase-space-slicing
method.\\
\\
{\em Typical running time:} Varies strongly from LO to NLO and depends
on type of subprocess (direct or resolved). At LO, running times of
about 1 minute for a cross section with fixed bin-size in one of the
kinematical variables are typical. At NLO, the running time for such a
cross section varies between 30 minutes (for the single resolved
contributions) to 5 hours (for the double resolved contributions).

\newpage

\input{main}
\newpage
\input{liter}

\newpage
\input{appendix}

\end{document}

%% file: main.tex
\section{Introduction}

One possibility to analyze hadronic final states in high energy
collision experiments is to cluster the final state particles into
jets with large transverse energy $E_T$, using a specific cluster
algorithm. In this way, inclusive single- and dijet cross sections can
be obtained for various physical processes. At HERA, jet cross
sections in electron-proton scattering have been experimentally
accessible for two distinct regions of the virtuality $Q^2$ of the
photon radiated by the lepton; one region is that of nearly on-shell
photons ($Q^2\simeq 0$) \cite{1,2,3}, the other is that of photons
with very large virtuality, $Q^2\gg 10$ GeV$^2$ \cite{4,5}. These two
regions define the photoproduction and the deep inelastic scattering
(DIS) regimes, respectively. On the theoretical side, perturbative QCD
calculations are possible due to the presence of a large scale in
these processes and single- and dijet cross sections have been
calculated in next-to-leading order (NLO) for both regimes. Two
different methods for these
higher order perturbative calculations have been used. In the phase
space slicing method, the singular phase space regions of soft and
collinear final state particles are separated by introducing an
invariant mass cut-off $y_s$. The finite phase space regions outside
the cut-off are calculated numerically. This method has been used in
photoproduction \cite{6,7,bourhis} and in DIS \cite{8,9}. The
subtraction method on the other hand is based on a point-by-point
subtraction of singularities in the numerical integration. This method
has also been applied in photoproduction \cite{10} and DIS \cite{11,12}.

Recently, the gap in $Q^2$ between photoproduction and DIS is being
closed by experiment and data becomes available for $eP$-scattering in
the region of intermediate photon virtuality, i.e., for $Q^2\simeq
[0,100]$ GeV$^2$ \cite{13Zeus, 13H1}. This has triggered theoretical
studies that try to match the photoproduction and DIS  regimes and to
provide a unified approach to treat virtual photons in a continuous
$Q^2$-range. From photoproduction it is well-known that the real
photon does not only interact directly with the partons from the
proton, but that it can also serve as a source of partons, i.e., quarks and 
gluons. These contributions are called {\em direct} and {\em resolved},
respectively. The resolved component is accompanied by a small $E_T$ 
remnant jet. In DIS, the resolved contribution should vanish. Thus,
one problem in matching the low and the high $Q^2$ regions is to
construct a parton distribution function (PDF) of the virtual photon,
which has been done by Gl\"uck, Reya and Stratmann (GRS) \cite{18} and
Schuler and Sj\"ostrand (SaS) \cite{19}. However, only limited data
exist for the virtual photon structure function; older measurements
have been performed by  the PLUTO collaboration at the $e^+e^-$
collider facility PETRA \cite{17}. Therefore, the available virtual
photon PDF's inhibit a rather larger uncertainty in shape and
magnitude, especially in the region of small $x$. Drees and Godbole
\cite{dg} have invented a simple $Q^2$-dependent interpolation factor
which multiplies the PDF's of the real photon.

Another theoretical question is how jet cross sections can be
calculated for all $Q^2$, which  has been addressed in a
number of works using leading order (LO) matrix elements
\cite{grs,14}. However, LO calculations suffer from large scale and scheme
dependences and are insensitive to any kind of jet clustering
algorithm. These problems can only be cured in NLO QCD. In NLO
photoproduction one has to subtract the initial state 
singularities arising from the photon to quark-antiquark splitting and
absorb them into the photon structure function, which introduces a
factorization scale dependence in the direct and resolved
components. The collinear parton from the photon splitting will
produce a low $E_T$ jet, similar to the remnant jet of the resolved
component. Therefore, the separation of the direct and resolved
processes is no longer possible in NLO. The results from
photoproduction can be extended to photons with moderate virtuality. 
The subtraction of photon initial state singularities for virtual photons
has been worked out using the phase space slicing method in \cite{15,16}. 
The subtraction term is absorbed into the $Q^2$-dependent 
virtual photon structure function. The NLO calculations for virtual
photons with direct and resolved components have been completed in \cite{20}
by including the longitudinally polarized direct photon contribution.

In the near future, measurements of jet cross sections from
$e\g$-scattering in a $Q^2$-range of the virtual photon similar to  
that at HERA will become available at LEP2 \cite{21,22} which
complements the HERA measurements. From the theoretical side, the
problems to be addressed in $e\g$-scattering are very similar to those
in $eP$-scattering, since the same types of subprocesses are
encountered. An extra contribution from the direct interaction of the
real and the virtual photon has to be considered. These contributions
have been worked out in \cite{16,eg}.

The goal of this paper is to describe how the calculations in
\cite{15,16,20,eg} are implemented in the computer package
\jv\footnote{\jv\ is an acronym for: {\bf Jet}s with {\bf Vi}rtual
  {\bf P}hotons.} for calculating jet cross sections in NLO QCD with
virtual photons. This program provides a link between the DIS
and the photoproduction regimes. The NLO calculations in deep-inelastic
$eP$-scattering have been implemented in several other computer
programs. The program  {\tt MEPJET}  \cite{8} uses the phase space
slicing method, whereas {\tt DISENT} \cite{11} and {\tt DISASTER++}
\cite{dis++} use the subtraction method. Two different strategies have
been applied in the programs for calculating jet cross sections. In
{\tt MEPJET}, a complete package is provided to handle the convolution
of the hard, perturbatively calculable partonic cross sections with
the parton density functions in the initial state and the
recombination of final state partons from the subprocess to jets. For
obtaining final jet cross sections, only a 
steering file has to be manipulated. In contrast, {\tt DISENT} and 
{\tt DISASTER++} generate an event of final state partons, weighted with
the respective  hard scattering cross section. The convolution over
the initial state and the recombination of the partons in the final
state into jets is left to the user. Of course, this allows a 
high flexibility, since any jet recombination or kinematical cut can
be implemented. The disadvantage is that the user will have to do a
lot of programming before jet cross sections can be calculated. 
A number of NLO programs exist also for the case of photoproduction in
$eP$- and $\g\g$-scattering \cite{6,7,bourhis,10}, but these have not
been published yet. 
It should be mentioned that cross checks of \jv\ with existing
programs in DIS and photoproduction have been performed which showed
good agreement \cite{24}.

For the sake of being user-friendly, in \jv\ we have adopted the strategy 
used in {\tt MEPJET}. All stages of the calculation are implemented in
the program. In particular we use the cone algorithm according to the 
Snowmass standard \cite{23} for combining partons in the final
state into jets with radius $R$. We use however the $R_{sep}$
parameter to modify the Snowmass algorithm to allow a higher
flexibility for comparison with experiments. Especially it is possible
to simulate $k_\perp$-like algorithms, containing an $R$ parameter. 
One feature of \jv\ is that the transverse energies and rapidities of the jets
are accessible so that cuts on these variables can be applied. By
choosing a zero cone radius, i.e., $R=0$, the transverse energies and
rapidities of the partons, rather than those of the jets, are
accessible and thus other jet recombination schemes than the two
mentioned here can be implemented by the user. Finally,
$\g^*\g$-scattering from $e^+e^-$ colliders can be studied  
with \jv. I will restrict myself in this paper to explaining the
computational techniques implemented in the package and the general
structure of the program. Physical applications have been discussed
elsewhere \cite{15,16,20,eg,24}. There, also the details of the NLO
calculations can be found. 

The outline of the paper is as follows. In section 2 the process of 
$eP$-scattering is discussed and the phase space slicing method is
explained. Details of the subtraction of the virtual photon initial state
contribution, which is vital for connecting the DIS and the
photoproduction limits, are given and the iterative cone algorithm is
discussed. In section 3 these results are extended to
photon-photon scattering. The input parameters of the steering file are
defined in section 4 and the range of the input parameters is
discussed. Last, section 5 contains a guide for the installation
of \jv\ on different computer platforms. Furthermore, typical running
times are given. The appendizes contain examples of the input- and
output-files, which can be used as simple cross checks after
installation of the program.

\section{Jet Cross Sections in Electron-Proton Scattering}

\subsection{General Structure of Cross Sections}

In order to define the general structure of the cross sections which
can be calculated with \jv, we write for the inclusive production of
two jets in electron-proton scattering
\equ{}{  e(k) + P(p) \rightarrow e(k')+ \mbox{jet}_1(E_{T_1},\eta_1) +
 \mbox{jet}_2(E_{T_2},\eta_2)+ \mbox{X} \quad . }
Here, $k$ and $p$ are the momenta of the incoming electron and proton,
respectively and $k'$ is the momentum of the outgoing electron. The
two jets in the final state are characterized by their transverse
momenta $E_{T_i}$ and rapidities $\eta_i$, which are the observables
also in the experiment. The four-momentum transfer of the electron is
$q = k-k'$ and $Q^2=-q^2$. The phase space of the electron is
parametrized by the invariants $y=pq/pk$ and $Q^2$. In \jv\ a
continuous range in $Q^2$ is covered. In the case of very small
virtualities $Q^2 \ll q_0^2$, where $q_0$  is the energy of 
the virtual photon, $y$ gives the momentum fraction of the initial
electron energy $k_0$, carried away by the virtual photon and
$y=q_0/k_0$. The total energy in the $eP$ center-of-mass system
(c.m.s.) is $\sqrt{S_H}$, where $S_H=(k+p)^2$. $W$ denotes the energy
in the virtual photon-proton ($\g^*P$) subsystem, $W^2=(q+p)^2$.

\begin{figure}[ttt]
\unitlength1mm
\begin{picture}(161,55)
\put(10,-14){\psfig{file=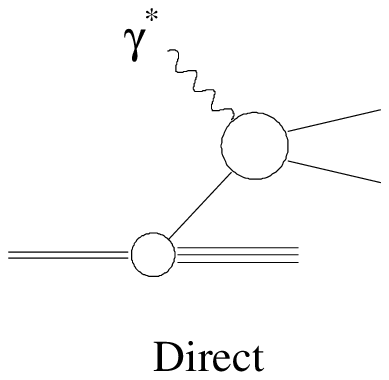,width=10.5cm}} 
\put(74,-14){\psfig{file=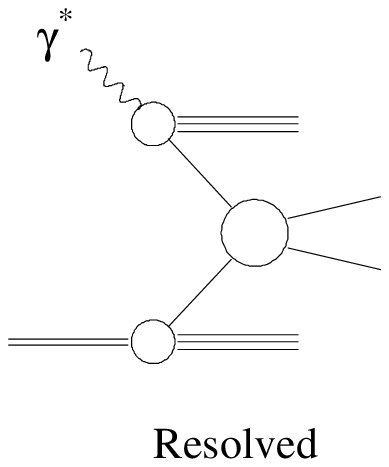,width=10.5cm}}
\put(0,8){\parbox[t]{161mm}{{\bf Figure 1:} \label{dir+res}The direct
     and resolved contributions in $\g^*P$-scattering.}}
\end{picture}
\end{figure}

As already mentioned in the introduction, in photoproduction, i.e.,
for $Q^2\simeq 0$, the photon can not only interact directly with the
partons from the proton, but can also have a non-perturbative,
resolved structure which has to be described by a structure
function. The concept of resolved photons can be extended to virtual
photons with moderate virtuality. In \jv, the direct and resolved
components for the virtual photon, shown in Fig.\ 1, are
implemented. In the following I will describe how cross sections for
the direct and resolved components in deep inelastic $eP$-scattering
are calculated with \jv, starting with the familiar direct case.

The hadronic cross section $d\si^H$ for the direct interaction is
written as a convolution of the hard 
scattering cross section $d\si_{eb}$, where the electron interacts
with the parton $b$ originating from the proton, parametrized by the PDF of
the proton $f_{b/P}(x_b)$ with $x_b$ denoting the parton momentum
fraction, so that 
\equ{}{ d\si^H(S_H) = \sum_{b}\int dx_b
  d\sigma_{eb}(x_bS_H)f_{b/P}(x_b) \quad . }
The cross section $d\si_{eb}$ for the scattering of the electron on
the parton $b$ is related to the lepton and hadron tensors
$L_{\mu\nu}$ and $H_{\mu\nu}$ by
\equ{}{ d\si_{eb} = \frac{1}{4S_Hx_b} \frac{4\pi\al}{Q^4}
  L^{\mu\nu}H_{\mu\nu} \ d\mbox{PS}^{(n)}\ dL \quad .}
Here,
\equ{}{  dL = \frac{Q^2}{16\pi^2} \frac{d\phi}{2\pi}
  \frac{dydQ^2}{Q^2} }
depends only on the electron variables, whereas $d\mbox{PS}^{(n)}$ depends
only on the $n$ final state particles from the hard interaction. The
azimuthal angle of the outgoing electron in the hadronic c.m.s.,
$\phi$, is integrated out in $d\si_{eb}$ with the result
\equ{xsec1}{ d\overline{\si}_{eb} = \int\limits_0^{2\pi}  d\phi \
  \frac{d\si_{eb}}{d\phi} = \frac{\alpha}{2\pi}\left(\frac{1+(1-y)^2}{y} 
  d\si^U_{\g b} + \frac{2(1-y)}{y}d\si^L_{\g b} \right)
  \frac{dydQ^2}{Q^2} \quad . }
In this formula the cross sections for the scattering of transversely 
unpolarized and longitudinally polarized virtual photons on the parton
$b$ are given by
\begin{eqnarray}
  d\si^U_{\g b} &=& \frac{1}{4x_bS_Hy}\ (H_g+H_L)\ d\mbox{PS}^{(n)} \quad , \\
 d\si^L_{\g b} &=& \frac{1}{2x_bS_Hy}\ H_L\ d\mbox{PS}^{(n)} \quad , 
\end{eqnarray}
with the definitions $H_g=-g^{\mu\nu}H_{\mu\nu}$ and
$H_L=\frac{4Q^2}{(S_Hy)^2}p^{\mu}p^{\nu}H_{\mu\nu}$. In the limit 
$Q^2\to 0$ one obtains the familiar formula for the 
absorption of photons with small virtuality, where $d\si^L_{\g b}$
is neglected and the transversely unpolarized cross section
$d\si^U_{\g b}$ is multiplied with the differential
Weizs\"acker-Williams spectrum 
\equ{ww11}{ \frac{df_{\g /e}}{dQ^2}
  =\frac{\alpha}{2\pi}\frac{1+(1-y)^2}{yQ^2}  \quad . }
When a photoproduction cross section is calculated with
\jv, where $Q^2\simeq 0$, formula (\ref{ww11}) is integrated over a small
$Q^2$-range, from $Q^2_{min}=m_e^2y^2/(1-y)$ up to a certain fixed
$Q^2_{max}$. For all other $Q^2>0$, the full formula (\ref{xsec1}) is
used instead of this approximation, in particular the longitudinal cross
section is included, which can account for up to 50\% of the full cross
section, depending on the kinematical conditions. The formula
(\ref{xsec1}) does not involve any approximations, except that terms
proportional to $m_e^2$ are neglected.

We now turn to the case, where a photon with moderate virtuality
interacts with the parton $b$ from the proton not as a point-like particle,
but via the partonic constituents of the photon. This partonic
structure of the photon is described by PDF's $f^{U,L}_{a/\g }(x_a)$,
introducing the new variable $x_a$ which gives the momentum fraction
of the parton in term of the virtual photon momentum,
$p_a=x_aq$. Since we must distinguish between transversely and
longitudinally polarized photons in (\ref{xsec1}), we must introduce two
PDF's for the photon with label U and L. To simplify the formalism we
can include the case of the direct photon interaction in the PDF's of
the photon by using  $f_{\g  /\g }^{U,L}=\delta(1-x_a)$ in the formula
below. Taking everything together, the hadronic cross section
$d\overline{\si}_H(S_H)$ can be written as a convolution of the hard
scattering cross section $d\si_{ab}$ for the reaction $a+b\rightarrow 
\mbox{jet}_1+\mbox{jet}_2+X$ with the PDF's of the photon
$f_{a/\g }^{U,L}(x_a)$ and the proton $f_{b/P}(x_b)$ in the following form
\equ{}{ \frac{d\overline{\si}_H(S_H)}{dQ^2dy}=\sum_{a,b} \int
  dx_adx_b\ f_{b/P}(x_b) d\overline{\si}_{ab}\ \frac{\alpha}{2\pi Q^2}
  \left[ \frac{1+(1-y)^2}{y} f^U_{a/\g}(x_a) +\frac{2(1-y)}{y}
    f^L_{a/\g}(x_a)  \right] \ . }
Of course, for the direct photon interaction $f^{U,L}_{a/\g }d\si_{ab}
= \de (1-x_a)d\si^{U,L}_{\g b}$. At this point we
mention that only the unpolarized PDF $f_{a/\g}^U$ is implemented in
\jv, for reasons that will become clear in section 2.3. For the
longitudinal PDF $f_{a/\g}^L$ only the delta function part is taken
into account. 

Since cross sections in \jv\ are calculated in the hadronic c.m.s.,
i.e., where the incoming photon and hadron are collinear, at least two
jets have to be present in the final state due to momentum
conservation. Thus, the LO cross section on the parton level has two
partons in the final state. The NLO corrections consist of the
one-loop contributions with two partons in the final state and the
real corrections with three partons in the final state. The different
kind of subprocesses implemented in \jv\ for the direct and resolved
cross sections are listed for the two-body final states in Tab.\
\ref{tLO}. These contain the LO Born contributions, if no internal
loops are present, and the NLO virtual corrections. The three-body
contributions are listed in Tab.\ \ref{tNLO} and are NLO. The direct
contributions contain both the transverse and the longitudinal photon
polarizations. The direct contributions implemented in \jv\ are taken
from Graudenz \cite{9}, the resolved contributions are taken from
Klasen, Kleinwort and Kramer \cite{6}. The $Z^0$-exchange, which is
negligible for $Q^2\ll M_Z^2$, is not implemented in \jv. In the
following section we explain how the NLO partonic cross sections are
calculated in principle with the phase space slicing method.

\begin{table}[ttt]
\renewcommand{\arraystretch}{1.6}
\caption{\label{tLO}Two-body subprocesses (Born and virtual)
  implemented in \jv.}
\begin{center}
\begin{tabular}{|c||c|c|c|} \hline
 direct & \multicolumn{3}{c|}{resolved} \\ \hline\hline 
 \makebox[2.9cm][c]{$\g^*q\to qg$} & 
 \makebox[2.9cm][c]{$qq'\to qq'$} & 
 \makebox[2.9cm][c]{$qq\to qq$} & 
 \makebox[2.9cm][c]{$qg\to qg$} \\ \hline  
 $\g^*g\to q\bar{q}$ & $q\bar{q}'\to q\bar{q}'$ & $q\bar{q}\to q\bar{q}$
 & $gg\to gg$ \\ \hline 
 & $q\bar{q}\to q'\bar{q}'$ & $q\bar{q}\to gg$ & $gg\to q\bar{q}$ \\ \hline
\end{tabular}
\end{center}
\end{table}

\begin{table}[ttt]
\renewcommand{\arraystretch}{1.6}
\caption{\label{tNLO}Three-body NLO subprocesses implemented in \jv.}
\begin{center}
\begin{tabular}{|c||c|c|c|} \hline
 direct & \multicolumn{3}{c|}{resolved} \\ \hline\hline 
 \makebox[2.9cm][c]{$\g^*q\to qgg$} & 
 \makebox[2.9cm][c]{$qq'\to qq'g$} & 
 \makebox[2.9cm][c]{$qq\to qqg$} & 
 \makebox[2.9cm][c]{$q\bar{q}\to ggg$} \\ \hline  
 $\g^*g\to q\bar{q}g$ & $q\bar{q}'\to q\bar{q}'g$ & 
 $q\bar{q}\to q\bar{q}g$ & $gg\to q\bar{q}g$ \\ \hline
 $\g^*q\to qq\bar{q}$ & $q\bar{q}\to q'\bar{q}'g$ & $qg\to qq\bar{q}$
 & $gg\to ggg$ \\ \hline  
 $\g^*q \to qq'\bar{q}'$ & $qg\to qq'\bar{q}'$ & $qg\to qgg$ & \\ \hline
\end{tabular}
\end{center}
\end{table}

\subsection{Partonic Cross Sections in NLO QCD}

The LO cross section $d\si^B$ on the parton level consists of the Born
matrix elements and a two-parton phase space. The NLO QCD corrections
to the LO cross section consist of the virtual corrections $d\si^V$
with a two-parton final state and the real corrections, where an
additional parton is radiated. Both these contributions have
characteristic divergencies.

The virtual corrections have infrared and ultraviolet singularities
due to the integration over the internal loops, which have been
calculated with the help of dimensional regularization. The
ultraviolet singularities are regularized in $d=4-2\e$ space-time
dimensions and subtracted in the modified minimal subtraction
($\overline{\mbox{MS}}$) scheme. The infrared divergences produce
$\e^{-n}$-terms, which become singular in the limit $\e\to 0$. These
cancel against poles in the real corrections.

The real corrections $d\si^R$ are singular in regions where one of the
outgoing partons is soft or collinear. These singular phase-space
regions are separated from the finite regions by applying the 
phase-space slicing method \cite{6}. To illustrate the method, we
write down the integration of over the phase-space of the third
particle, somewhat symbolically, as 
\equ{pps}{ d\si^R = \int\limits_0^1 d\mbox{PS}^{(3)} 
  |{\cal M}_{2\to 3}|^2 = \int\limits_0^{y_s} d\mbox{PS}^{(3)} 
  |{\cal M}_{2\to 3}|^2 + \int\limits_{y_s}^1 d\mbox{PS}^{(3)} 
  |{\cal M}_{2\to 3}|^2 }
where we have inserted a technical slicing parameter $y_s$ to separate
different regions of phase space. The first integral on the rhs of this
equation, which we denote as $d\si_2(y_s)$, contains the singular parts,
where one of the final state particles becomes soft or collinear and
can not be observed. To integrate the singular piece $d\si_2$, the
three-body phase space is split into a two-body phase space (of the
two remaining particles) and a singular phase space
$d\mbox{PS}^{(s)}$, which is integrated out using dimensional
regularization, as for the virtual corrections. Effectively a
two-parton final state remains:
\equ{}{ d\si_2(y_s) = d\mbox{PS}^{(2)} \int\limits_0^{y_s}
  d\mbox{PS}^{(s)} |{\cal M}_{2\to 3}|^2 \quad . }
The second integral, which we denote as $d\si_3(y_s)$, is finite due
to the cut-off at the lower integration boundary. \jv\ integrates out
$d\si_3$ numerically using the Monte-Carlo integration package
{\tt VEGAS} \cite{25} to handle the multi-dimensional phase-space
integrals. In the numerical part, a variety of experimental cuts and
jet definitions can be implemented. This flexibility allows a detailed
comparison between theory and experiment. The $y_s$-dependence of
$d\si_3(y_s)$ has to cancel against that of the two-body
contributions $d\si_2(y_s)$, so that the NLO real corrections,
$d\si^R= d\si_2(y_s)+d\si_3(y_s)$, are independent of $y_s$. In the
actual calculation of $d\si_2$, contributions of order 
${\cal O}(y_s\ln^ny_s)$ are neglected, so $y_s$ has to be chosen
sufficiently small in the numerical part $d\si_3$ for the cancellation
to be valid.

The two-body part of the real corrections can be further distinguished
into initial state and final state corrections, $d\si_2 = d\si^I +
d\si^F$, where a particle in the initial or final state produces a
singularity. The infrared poles in $d\si_2$ are canceled by poles
from the virtual corrections, except for simple poles in $\e$ from the
initial state which are multiplied by a splitting function 
$P_{a\leftarrow b}$. The singular terms are absorbed into the scale dependent
structure function of the hadron $H$ the initial parton was emitted from:
\equ{rpdf}{  f_{a/H}(\eta,M_F^2) = \int\limits_\eta^1 
  \frac{dz}{z} \left(\de_{ab}\de (1-z) + 
    \frac{\al_s}{2\pi} \G_{a\leftarrow b}^{(1)}(z,M_F^2) \right)
  f_{b/H}\left(\frac{\eta}{z}\right) \quad , }
where $f_{b/H}$ is the LO PDF before the absorption of the collinear
singularity and $M_F^2$ is the factorization scale. 
The transition function $\G^{(1)}_{a\leftarrow b}(z)$ contains the pole term
and some finite part $C_{a/b}$, which is chosen to be zero in the 
$\overline{\mbox{MS}}$ scheme:
\equ{transf}{ \G^{(1)}_{a\leftarrow b}(M_F^2)(z) = -\frac1\e
  P_{a\leftarrow b}(z) \left( \frac{4\pi\mu^2}{M_F^2}\right)^\e 
        \frac{\G (1-\e )}{\G (1-2\e )} + C_{a/b}(z) \quad . } 
The $\e$-dependent factors are artefacts of the dimensional
regularization. The renormalized initial state cross section $d\si^I$ is
calculated from the unrenormalized cross section $d\bar{\si}^I$ by
subtracting the singular terms convoluted with the Born cross section:
\equ{rini}{ d\si^I_{ab} = d\bar{\si}_{ab} - \frac{\al_s}{2\pi} \int dz
  \G_{b\leftarrow b'}^{(1)}(z,M_F^2) d\si^B_{ab'}
  \quad .}
After this subtraction procedure, the sum of all LO and NLO two-body
contributions 
\equ{}{ d\si^B + d\si^V + d\si_2(y_s) =
  d\si^B + d\si^V + d\si^I(y_s) + d\si^F(y_s) }
is finite and the limit $\e\to 0$ can be safely performed. Adding the
two-body contributions and the three-body contributions
$d\si_3(y_s)$ then yields a physically meaningful NLO cross section,
which is finite and independent of $y_s$. 

Since the subtraction of initial state contributions for the virtual
photon is of special interest for the transition from photoproduction
to DIS, I will explain this point in more detail in the following
section.

\subsection{Subtraction of Virtual Photon Initial State Contributions}

In \jv\ two different concepts for calculating jet cross sections in
$eP$-scattering for finite $Q^2$ are realized:
\begin{enumerate}
\item The virtual photon couples directly to the
  partons from the proton, which reflects the classical situation
  known as deep-inelastic scattering. The integration of the NLO
  matrix elements over the three-body final state is not truncated
  on the virtual photon side at a small cut-off $y_s$, since all
  contributions are finite. They may, however, be large for $Q^2\ll
  E_T^2$. 
\item Large logarithms from the virtual photon side (for $Q^2\ll E_T^2$) 
  are separated from contributions where the photon couples directly
  to the partons from the proton by introducing a lower integration
  boundary $y_s$ in the phase space integrations of the three-body
  final state. These logarithms are absorbed into the virtual photon
  structure function. Therefore one has to add a resolved
  contribution, where the photon serves as a source of partons that
  interact with the partons from the proton. 
\end{enumerate}
Since it is a priory not clear up to which virtualities the concept of
a resolved virtual photon makes sense, always both approaches should
be considered for comparison. Normally, this would require two
separate calculations, one for each approach, since the phase space
integrations are truncated in the numerical part for the
second approach. However, as will become clear in the following, the
subtraction term can be calculated separately, so that no additional
CPU time is required.

We start by reminding that the real, i.e., massless, photon can decay
into a quark-antiquark pair in the initial state, which becomes
collinear and produces a singularity. The subtraction of photon
initial state singularities for the real photon has been worked out in
the phase space slicing method in \cite{6}. The procedure for
subtracting the contributions from the initial state virtual photon
is completely analogous \cite{15,16}. One calculates the $2\to 3$
matrix elements with $Q^2\ne 0$ and decomposes them into terms with
the characteristic denominator from $\g^*\to q\bar{q}$ splitting which
become singular in the limit $Q^2\to 0$. 
The singular terms for $Q^2\to 0$ of the matrix elements for
transversely polarized photons after phase space integration up to a
cut-off $y_s$ have the same structure as in the real photon case. The
integration can be done with $\e = 0$  since $Q^2\ne 0$. In the real
photon case the pole term has the form $\frac1\e
P_{q\leftarrow\g}(z)$, where $P_{q\leftarrow\g}(z)$ is the photon to 
quark splitting function.  Instead of the pole, for the virtual photon
the logarithm  
\equ{}{ M = \ln\left( 1 + \frac{y_ss}{zQ^2}\right)
  P_{q\leftarrow\g}(z) } 
occurs, which is singular for $Q^2 = 0$. This singularity is absorbed
into the PDF of the virtual photon with virtuality $Q^2$. The
transition function (compare eqn (\ref{transf})) reads
\equ{}{ \G^{(1)}_{q\leftarrow\g}(z,M_\g^2) =
  P_{q\leftarrow\g}(z) \left[ \ln\left( \frac{M_\g^2}{Q^2}\right)  -
    \ln (1-z) \right] - 1 \quad . }  
After absorbing $\G^{(1)}_{q\leftarrow\g}$ into the virtual photon PDF
in analogy to eqn (\ref{rpdf}), and subtracting the logarithm from
the unrenormalized cross section in analogy to eqn (\ref{rini}), the 
remaining finite term (for $Q^2\to 0$) in $M$ yields 
\equ{12}{ M(Q^2)_{\overline{MS}} = - P_{q\leftarrow\g}(z) \left[
    \ln\left(\frac{M_\g^2z}{zQ^2+y_ss}\right) - \ln (1-z) \right] 
  + 1 \quad . }
In addition to the singular term  $\ln (M_\g^2/Q^2)$, two finite terms
have been subtracted in order to achieve the same result as for the
$\overline{\mbox{MS}}$ factorization in the real photon case, when the
limit $Q^2\to 0$ is taken in (\ref{12}). Therefore this form of
factorization is called the $\overline{\mbox{MS}}$ factorization for
$Q^2\ne 0$. The term (\ref{12}) is the one implemented in \jv\ for
calculating the virtual photon initial state contribution. 

Taking a close look at eqn (\ref{12}) one observes that it is not
necessary to have a non-zero cut-off $y_s$ to calculate the
subtraction term. As long as $Q^2\ne 0$, it will suffice to choose
$y_s=0$ for evaluating the leading logarithmic contribution from the
$\g^*\to q\bar{q}$ splitting. This has a numerical advantage for the
comparison of the two approaches mentioned at the beginning of this
section. Since $y_s=0$, no cut-off is introduced in the phase space
integration on the virtual photon side for the three-body direct
contributions, which gives the DIS result. Thus, for comparing the two
approaches, only three calculations have to be performed. After
calculating the (time-consuming) DIS cross section, where $y_s=0$ on the
virtual photon side, one can calculate the subtraction term (\ref{12})
with $y_s=0$, which numerically can be done very fast. Third, the
resolved cross section is calculated. The approach 1 is then given by
the DIS result with $y_s=0$, the approach 2 is given by adding the
resolved cross section and the {\em subtracted direct} cross section,
which is obtained by adding the DIS result and the subtraction term.

However, this method can only be applied as long as $Q^2$ is not too
small. If $Q^2$ becomes too small, the DIS cross section will start to
become too large due to the logarithms $\ln (Q^2/E_T^2)$ and the numerical
integration becomes unstable and produces large statistical errors. 
Thus, one has to choose a finite $y_s$ in the numerical phase space
integration and in the subtraction term (\ref{12}). I have observed
that setting $y_s=0$ works well for all $Q^2\ge 1$ GeV$^2$. Below
$Q^2=1$ GeV$^2$ the photoproduction regime begins and it is definitely
necessary to take into account the resolved component, i.e., only
approach 2 will give appropriate results. Then one has to choose a
finite $y_s$. 

In \jv\ only the photon splitting terms for the transversely polarized
photons are implemented. The longitudinal parts are proportional to
$Q^2$ and vanish in the limit $Q^2\to 0$. Since they do not produce a
singularity, one does not necessarily have to introduce a longitudinal
virtual photon structure function. Apart from this argument, no
longitudinal virtual photon structure function has been constructed so
far.

\subsection{The PDF of the Virtual Photon}

The scale ($\mu^2$) dependence of the parton distributions of the
virtual photons with virtuality $Q^2\ne 0$ in the region $\Lambda^2\ll
Q^2\ll \mu^2$, with $\Lambda^2$ being a soft scale, follows from the
respective evolution equations \cite{uw81}. These are very
similar to those for the real photon case with $Q^2=0$ and contain a
point-like and a hadronic part. Of special interest is the region
where  $Q^2\simeq \Lambda^2$. 

As mentioned in the introduction, mainly two different LO
parametrizations of the virtual photon PDF have been constructed. These 
are implemented in \jv. The construction is done such that the case
of the real photons is reproduced for $Q^2\to 0$ and the exact
evolution equations are obeyed for $Q^2\gg \Lambda^2$:
\begin{itemize}
\item Gl\"uck, Reya and Stratmann (GRS) \cite{grs}: The evolution
  equations are solved with a smooth interpolation of the boundary
  conditions valid at $Q^2=0$ and for $Q^2\gg \Lambda^2$. LO and NLO
  parton distributions have been obtained in this way, but only the LO
  ones are parametrized in a form which is convenient for numerical
  calculations.

  The application of the GRS PDF's is restricted to
  $Q^2\le 10$ GeV$^2$, $\mu^2\in [0.6,5\cdot 10^4]$ GeV$^2$ and
  $Q^2\le 5\mu^2$. Furthermore, $x\in [10^{-4},1]$. The number of
  flavours is limited to 3. 

\item Schuler and Sj\"ostrand (SaS) \cite{19}: The dipole integral over the
  virtuality $k^2\in [0,\mu^2]$ of the $\g^*\to q\bar{q}$ state is
  modified by the factor $(k^2/(k^2+Q^2))^2$. The integration over the
  low $k^2$ region, $k^2<\mu_0^2$, is associated with the
  hadronic part, whereas the high $k^2$ region, $k^2>\mu_0^2$, is
  associated with the pointlike part. Different values of $\mu_0^2$
  can be selected.

  The application of the SaS
  PDF's is not restricted kinematically, although it should be noted
  that in the region $Q^2>\mu^2$ the PDF's are nearly zero. 
\end{itemize}
As a third possibility, in \jv\ the Drees and Godbole
interpolation factor is implemented:
\begin{itemize}
\item Drees and Godbole (DG) \cite{dg}: The quark distributions of the
  real photon are multiplied with the scaling factor 
  \[ r = 1 - \frac{\ln (1+Q^2/P_C^2)}{\ln (1+\mu^2/P_C^2)} \quad , \]
  where $P_C^2$ is a typical hadronic scale of order $0.5$ GeV$^2$
  (the value $P_C^2=0.5$ GeV$^2$ is fixed in \jv). The stronger
  suppression of the gluon distribution is modelled by multiplying
  these distributions with $r^2$ instead of $r$. 

  For $Q^2=0$, $r=1$ (real photon case). For $Q^2=\mu^2$, $r=0$ (no
  resolved contribution). Note, that $r<0$ for $Q^2>\mu^2$. 
\end{itemize}
In lepto-production $E_T^2$, $Q^2$ or a combination of both are
possible choices for the hard scale $\mu^2$. In the transition region,
the combination $\mu^2=E_T^2+Q^2$ reproduces the scale $E_T^2$, used
in photoproduction, in the limit $Q^2\to 0$. In the DIS regime for
$Q^2\gg E_T^2$ one has $\mu^2\simeq Q^2$. Obviously, as can be seen
from the restrictions of the PDF's of GRS and SaS, the choice of the
scale will dramatically affect whether a resolved component is present
in the theoretical calculations or not, especially in a region where
$Q^2\simeq E_T^2$. This is true also for the DG
model. To obtain a smooth transition from the
photoproduction to the DIS region, without having to bother about
discontinuities on the PDF's of the resolved photon, one can choose
the SaS parametrization with the scale $\mu^2=E_T^2+Q^2$, since here
the restriction $Q^2\le 5\mu^2$ is not built in as for the GRS
parametrization. This model dependence however only reflects the fact
that the structure and especially the $Q^2$-dependence of the virtual
photon is not known well. Of course, if new parametrizations are
released in the future, these will be included in \jv.

It is clear that in NLO the PDF for the virtual photon must be
given in the same scheme that has been used for the calculation of the
$\g^*\to q\bar{q}$ splitting term, i.e., the $\overline{\mbox{MS}}$
factorization scheme. In \cite{18} the PDF is constructed in the
DIS$_\g$ scheme, which is defined as for real photons ($Q^2=0$). This
distribution function is related to the $\overline{\mbox{MS}}$ scheme
PDF in the following way \cite{18}:
\equ{}{ f_{a/\g}(x,M_\g^2)_{DIS_\g} = 
  f_{a/\g}(x,M_\g^2)_{\overline{MS}} + 
  \de f_{a/\g}(x,M_\g^2) }
where
\equ{14}{ \de f_{q_i/\g}(x,M_\g^2)
  = \frac{\al}{2\pi} \bigg[ P_{q_i\leftarrow\g}(x)   \ln\left(
    \frac{1-x}{x} \right) +  8x(1-x) - 1 \bigg] \quad . }
Furthermore, $\de f_{q_i/\g} = \de f_{\bar{q}_i/\g}$ and 
$\de f_{g/\g}=0$. If the PDF in this scheme is used to calculate the
resolved cross section one must transform the NLO finite terms in the
direct cross section. This produces a shift of $M(Q^2)_{\overline{MS}}$ 
as given in (\ref{12}) by the same expression as in (\ref{14}). The
appropriate expression has been implemented in \jv\ to account for the
DIS$_\g$ scheme. Strictly speaking, the transformation (\ref{14}) only
makes sense for NLO PDF's. Since these are not available yet, we treat
the LO parametrizations as if they were NLO. Note, that SaS have defined a
slightly different DIS$_\g$ scheme than GRS. The appropriate
transformation formul{\ae} for going from one scheme to the other for
the SaS PDF's are not implemented in \jv. However, using eqn (\ref{14}) will 
produce results which are very similar to the correct transformation.

\subsection{Jet Definitions}

In the following I will discuss two of the main jet algorithms which
are used in experiments. One is the iterative cone algorithm
\cite{cone}, the other is the $k_\perp$ algorithm \cite{kt}. In both
cases, a jet is defined in terms of its constituent particles. According to
the Snowmass convention \cite{23}, the transverse energy $E_{T_J}$,
pseudorapidity $\eta_J$ and azimuth angle $\phi_J$ of a
jet are defined as 
\begin{eqnarray}
 E_{T_J} &=& \sum_i E_{T_i} \quad , \nonumber \\
 \eta_J  &=& \frac{1}{E_{T_J}} \sum_i E_{T_i}\eta_i \quad , \label{jets}
 \\
 \phi_J  &=& \frac{1}{E_{T_J}} \sum_i E_{T_i}\phi_i \quad \nonumber ,
\end{eqnarray}
where the sum runs over all partons inside the jet. The
(boost-invariant) opening angle between two partons $p_i$ and $p_j$ 
in the $(\eta,\phi)$-plane is defined as 
$R_{ij}=\sqrt{(\eta_i-\eta_j)^2+(\phi_i-\phi_j)^2}$.

The iterative cone algorithms, as e.g.\ used by the CDF and D0
collaborations, consist of the following steps \cite{seym1}: 
\begin{enumerate}
\item After particles have been clustered into calorimeter cells (CC) of
  certain size in $(\eta,\phi)$-space, each CC above a certain energy
  $E_0$ is considered as a {\em seed cell} with direction
  $(\eta_S,\phi_S)$.

\item For each seed cell, jets are defined by summing all CC $i$
  according to eqn (\ref{jets}), if $R_{iS}<R$ with a certain cone radius
  $R$. If the jet direction does not coincide with the seed direction,
  the seed direction is replaced by the jet direction. This step is
  iterated, until a stable jet direction is found. 

\item All jet duplicates are thrown away. 
\end{enumerate}
Finally, jets which are overlapping have to be taken care of. For
these, a splitting procedure is defined, which is slightly different
in different experiments. In principle, jets are merged which have a
certain percentage of their energy in common (typically 50\% to 75\%)
with the direction given by the higher-energy jet, and split otherwise. 

In \jv, partons are combined into a jet with direction
$(\eta_J,\phi_J)$ also according to eqn (\ref{jets}). A parton $p_i$ is
included into the jet, if the condition
\equ{}{ \sqrt{(\eta_i-\eta_J)^2+(\phi_i-\phi_J)^2}<R}
is fulfilled, where $R$ is the same as in step 2 in the above
procedure. This definition is analogous to combining two partons $p_i$
and $p_j$ into a single jet, if they fulfill the condition
\equ{jcond}{ R_{ij} \le R\cdot E_{ij} \qquad \mbox{with} \qquad 
 E_{ij} \equiv \frac{E_{T_i}+E_{T_j}}{\mbox{max}(E_{T_i},E_{T_j})} \quad . }
This kind of parton level jet definition is very similar to the 
steps 1--3 in the above described algorithm. Since at
NLO we have at most three partons in the final state, the theoretical
procedure is of course much simpler. There are no iterations needed to
define stable jets and at maximum only two partons can be merged in the
hadronic c.m.s.\ due to momentum conservation. Finally, no merging or
splitting of two clusters depending on their shared energy occurs.

There is, however, a well-known problem for iterative cone algorithms
\cite{seym1}. It arises from configurations, where 
particles with balanced $E_T$ have a distance between $R$ and $2R$, so
that both particles are within $R$ of their common center, see Fig.\ 2
(a). For this configuration, the iterative cone algorithm will
produce two stable jets since no other particle is in the intermediate
region.

This situation does not change if an additional soft particle below
the threshold $E_0$ is added in the center between particles $p_i$ and
$p_j$, see Fig.\ 2 (b). However, if the third particle is slightly
above the threshold, it will be considered as an additional seed
cell. Since the cone around the seed encloses the other two particles,
all three particles will be merged into a single stable jet, in
contrast to the two-jet configuration without an additional soft
particle. It is clear that the so found cross sections depends on the
energy threshold $E_0$. Thus, the iterative cone algorithm is not
fully infrared safe. Furthermore, if the soft particle above $E_0$
splits into two collinear particles slightly below $E_0$ in two
slightly different CC, the single-jet configuration will flip back to
a two-jet one. So the iterative cone algorithm is also not fully
collinear safe. 

\begin{figure}[ttt]
\unitlength1mm
\begin{picture}(161,70)
\put(-5,3){\psfig{file=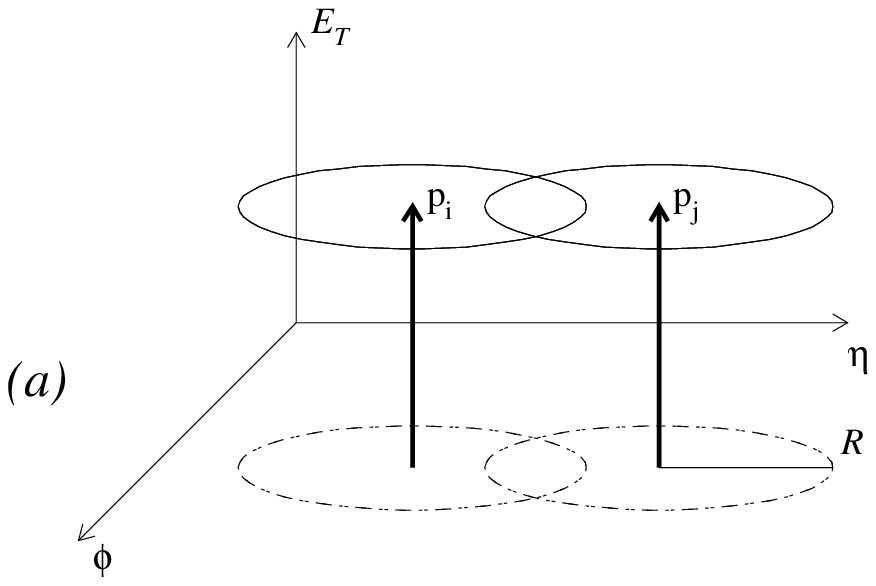,width=8.4cm,height=9.6cm}} 
\put(74,3){\psfig{file=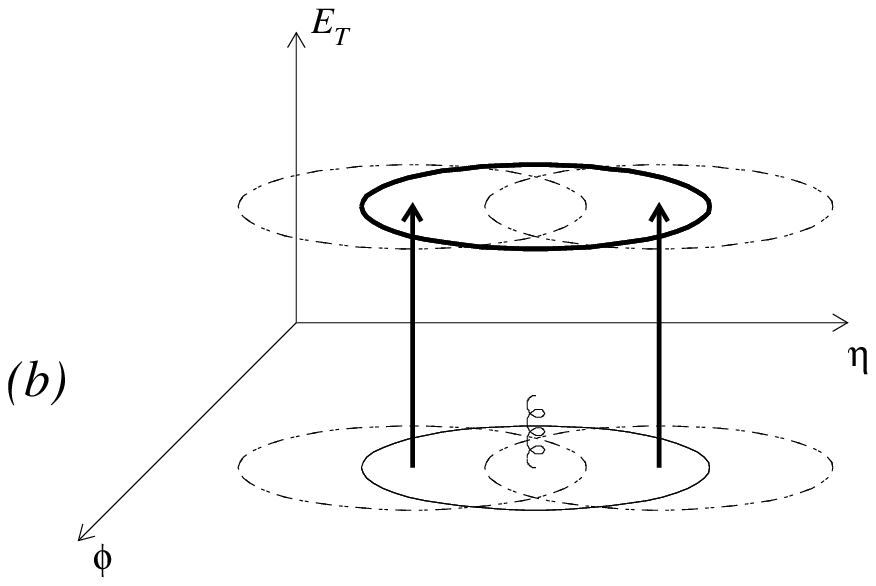,width=8.4cm,height=9.6cm}}
\put(0,12){\parbox[t]{161mm}{{\bf Figure 2:} (a) Two partons
    $p_i$ and $p_j$ in the $(\eta,\phi)$-plane with a distance between $R$
    and $2R$ with balanced $E_T$. (b) Additional soft particle in the
    overlap region (figure adapted from \cite{seym1}).}}
\end{picture}
\end{figure}

At NLO, this behaviour can not be observed, since
there are no soft particles available. To simulate the overlap problem
in the theoretical analysis in a simple way, Ellis, Kunszt and Soper have
introduced the concept of the phenomenological $R_{sep}$ parameter 
\cite{eks}. EKS suggested to imply the constraint, that the distance
between the two partons should not only fulfill eqn (\ref{jcond}),
but also be less than $R_{sep}$. Thus, two partons $p_i$ and $p_j$ are
combined into a single jet, if the condition $R_{ij}\le
\mbox{min}(R\cdot E_{ij},R_{sep})$ is fulfilled. 

A number of objections can be raised against this kind of solution,
which have been discussed in \cite{seym1}. A jet definition that
avoids all the discussed problems is the $k_\perp$
algorithm. Including an $R$ parameter the steps in the experimental
analysis are \cite{kt}:
\begin{enumerate}
\item For every pair of particles, define a closeness
  $d_{ij} = \mbox{min}(E_{T_i},E_{T_j})^2R_{ij}^2$
  (for small opening angles $R_{ij}\ll 1$, one finds 
  $d_{ij}\simeq \mbox{min}(E_i,E_j)^2\theta_{ij}^2\simeq k_\perp^2$).
\item For every particle define a closeness to the beam particles
  $d_{iB}=E_{T_i}^2R^2$. 
\item If $\mbox{min}(d_{ij})<\mbox{min}(d_{iB})$, merge particles
  $p_i$ and $p_j$ according to eqn (\ref{jets}). If otherwise
  $\mbox{min}(d_{ij})>\mbox{min}(d_{iB})$, jet $i$ is complete. 
\end{enumerate}
These steps are iterated until all jets are complete. As a result, all
jets are at least $R$ apart and all opening angles within each jet are
less than $R$. The NLO calculation produces jets equal to
those found in the $k_\perp$ algorithm by combining two particles
$p_i$ and $p_j$ according to eqn (\ref{jets}), if they 
fulfill the condition $R_{ij}<R$, that is by choosing the $R_{sep}$
modified cone algorithm with $R_{sep}=R$ (since $E_{ij}\ge 1$).

\section{Jet Cross Sections in Photon-Photon Scattering}

We will now extend the case of $eP$-scattering to include the
scattering of virtual on real photons, as it can be achieved at
$e^+e^-$ colliders. 

\subsection{General Structure of Cross Sections}

To fix the notation we start by writing down the process of 
jet production in $e^+e^-$ scattering:
\equ{}{ e^+(k_a) + e^-(k_b) \longrightarrow e^+(k_a') + e^-(k_b') +
  \mbox{jet}_1(E_{T_1},\eta_1) +\mbox{jet}_2(E_{T_2},\eta_2) +
  \mbox{X}  \quad .} 
We are interested in the case where one lepton radiates a virtual and
the other a real photon. Of course, it does not matter which of the
leptons radiates the virtual photon, but for definiteness we suppose
this to be the positron. Thus, the subprocess we have to consider is 
$\g^*_a(q_a) + \g_b(q_b) \to \mbox{jet}_1 + \mbox{jet}_2 + \mbox{X}$,
with $q_a = k_a-k_a'$, $q_b = k_b-k_b'$ and the virtualities $Q^2 =
-q_a^2$ and $P^2 = -q_b^2=0$. The electron-positron center-of-mass
energy is $s_H=(k_a+k_b)^2$. The energy in the hadronic, i.e., $\g^*\g$,
c.m.s.\ is $W^2 = (q_a+q_b)^2$. Furthermore we define the
variables 
\equ{}{ y_a = \frac{q_ak_b}{k_ak_b} \qquad \mbox{and}  \qquad  
  y_b = \frac{q_bk_a}{k_ak_b} \simeq 1-\frac{E_e'}{E_e} \quad , } 
where $E_e$ and $E_e'$ are the energies of the incoming and outgoing
electron in the $e^+e^-$ c.m.s., respectively. The variable $y_b$
gives the momentum fraction of the real photon in the electron.

As for $eP$-scattering, the virtual photon can have a point-like part
and a non-perturbative, resolved structure for moderate $Q^2$. The
point-like and hadronic components are likewise found for the real
photon with $P^2=0$. Taking all possible interaction modes into
account, one finds in total four contributions for the interaction of
real and virtual photons (see Fig.~3):
\begin{enumerate}
\item Direct (D): both photons couple directly to the charge of the
  quarks.
\item Single-resolved (SR): the real photon is resolved and the
  virtual photon interacts directly with the partons from the
  photon.
\item Single-virtual-resolved (SR$^*$): the virtual photon is resolved
  and the real photon interacts directly with the partons from the
  virtual photon.
\item Double-resolved (DR): both photons are resolved.
\end{enumerate}
The contributions 2 and 4 are familiar from $eP$-scattering with the
resolved photon substituted by a proton and thus have a structure
very similar to that encountered in DIS. The subprocesses for these
contributions are listed in Tab.\ 1 and 2.
The subprocesses of contribution 3 are easily obtained from
contribution 2 by setting $Q^2=0$. Contribution 1 has no counterpart
in $eP$-scattering and has been calculated in NLO QCD in \cite{eg}. It
consists in LO of the Born process $\g^*\g\to q\bar{q}$ and in NLO of the
one-loop corrections to the Born process and the real gluon emission
process $\g^*\g\to q\bar{q}g$. This completes the list of processes
implemented in \jv as given in Tab.\ 1 and 2. 

\begin{figure}[ttt]
\unitlength1mm
\begin{picture}(161,54)
\put(-30,-12){\psfig{file=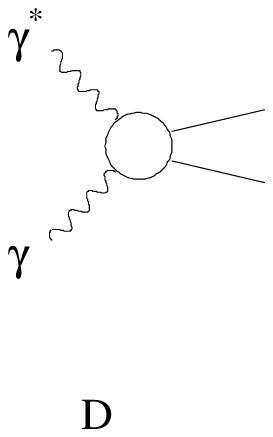,width=10.4cm}}
\put(10,-12){\psfig{file=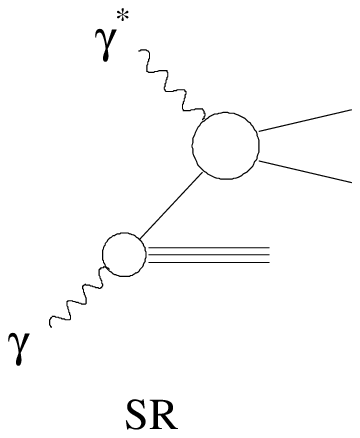,width=10.4cm}}
\put(50,-12){\psfig{file=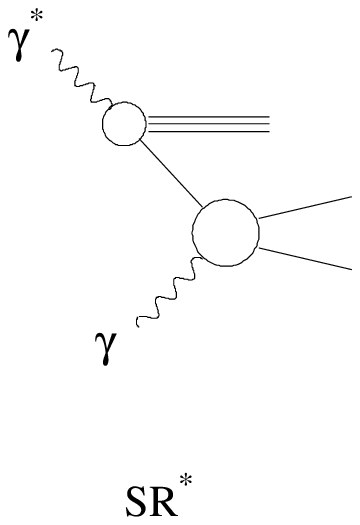,width=10.4cm}}
\put(90,-12){\psfig{file=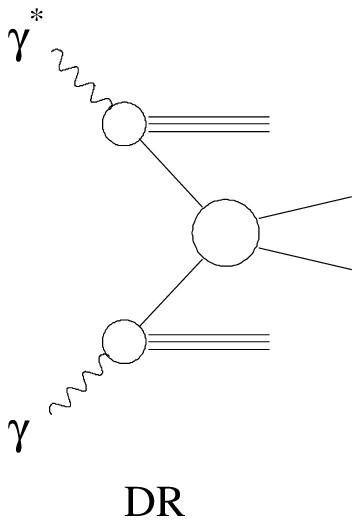,width=10.4cm}}
\put(0,2){\parbox[t]{161mm}{{\bf Figure 3:} The
    different components contributing in $\g^*\g$ scattering.}}
\end{picture}
\end{figure}

Taking into account both the transverse and longitudinal polarizations
of the virtual photon, the cross section $d\si_{e^+e^-}$ for the
processes 1--4 is conveniently written as the convolution
\begin{eqnarray} 
  \frac{d\si_{e^+e^-}}{dQ^2dy_ady_b} &=& \sum_{a,b} \int dx_adx_b \
  F_{\g /e^-}(y_b) \ f_{b/\g}(x_b) \  d\si_{ab} \nonumber \\
 &\times& \frac{\al}{2\pi Q^2} \left[ \frac{1+(1-y_a)^2}{y_a}
  f^U_{a/\g^*}(x_a) +\frac{2(1-y_a)}{y_a} f^L_{a/\g^*}(x_a)  \right]
  \quad . \label{e+e-}  
\end{eqnarray}
The PDF's of the real and the virtual photon are $f_{b/\g}(x_b)$ and 
$f^{U,L}_{a/\g^*}(x_a)$, respectively. The direct processes are
included in formula (\ref{e+e-}) through delta functions. For the
direct virtual photon one has 
$f^{U,L}_{a/\g^*}d\si_{ab} = \de (1-x_a)d\si^{U,L}_{\g^*b}$, 
whereas for the direct real photon the relation is 
$f_{a/\g}d\si_{ab} = \de (1-x_b)d\si_{\g b}$. As for the $eP$ process,
only the unpolarized PDF $f_{a/\g}^U$ is implemented in
\jv, because the subtraction procedure has only been worked out for
that case. 

The function $F_{\g /e^-}(y_b)$ describes the spectrum of the real photons
emitted from the electron according to the Weizs\"acker-Williams
approximation \cite{wwill}, which in its' simplest form reads
\equ{ww}{ F_{\g /e^-}(y_b) = \frac{\al}{2\pi} \frac{1+(1-y_b)^2}{y_b}
  \ln \left( \frac{E_e^2\theta^2_{max}}{m_e^2} \right) \quad .}
The electron mass is $m_e$ and $\theta_{max}$ is the maximum
scattering angle of the electron. 

All other aspects of jet cross sections that have been addressed in the
connection with $eP$-scattering, as there are the subtraction of
virtual photon initial state singularities, the virtual photon
structure function, or the jet algorithms, remain unchanged for
$\g^*\g$ reactions. The frame of reference in which cross sections are
calculated is the hadronic (i.e., $\g^*\g$) c.m.s.

\subsection{Notation of Subprocesses in \jv}

As we have seen, the $\g^*\g$ reactions and the $eP$ processes have
quite a number of subprocesses in common. To unify the notation and to
therefore simplify the handling of the input-file, which will be
described in the following section, we will denote the direct process
in $eP$-scattering as SR, whereas the resolved contribution is denoted
as DR.

\section{Definition and Range of Input Parameters}

Now that the theoretical concepts underlying the program have been
clarified, we turn our attention to the details of setting
the parameters in the input-file to obtain a jet cross section from
\jv. The input-file needs to have the exact format as shown in
appendix A (the file shown in the appendix contains a set of standard
values to produce a single-jet inclusive LO cross section in $eP$-scattering
for HERA conditions). The input-file has five main categories, which are:
\begin{enumerate}
\item Specification of the process and contributions to the process.
\item Specification of the initial state (of the real and virtual
  photons and/or proton). This includes the energies of
  the incoming particles and the kinematics of the incoming and
  outgoing leptons.
\item Specification of the kinematics and parameters relevant for 
  calculating the subprocess. This includes the specification of the
  parton densities for the photon and/or proton and of the scales. 
\item Specification of the final state jets. 
\item Specification of the parameters for the {\tt VEGAS} integration
  routine.
\end{enumerate}
In the following, the definition and range of the parameters occurring
in these categories will be explained. Numbers in squared brackets
point to the section where the relevant theoretical concept is
discussed.

\subsection{Contribution}

\begin{itemize}
\item {\tt iproc}: integer $\in \{ 1,2\}$. Type of process.
  $1=eP$, $2=e^+e^-$
\item {\tt ijet}: integer $\in \{ 1,2\}$. $1=$ single-jet, $2=$ dijet.
\item {\tt isdr}: integer $\in \{ 1,2,3,4\}$. Selection of
  the D, SR, SR$^*$ or DR contribution. [sec 3.2] $1=$D; $2=$SR;
  $3=$SR$^*$; $4=$DR. Only one of the contributions D, SR, SR$^*$ or
  DR can be calculated at a time. They have to be added after the
  calculation.
\end{itemize}
In the following, the LO and NLO contributions are turned on ($1$) or
off ($0$):
\begin{itemize}
\item {\tt iborn}: integer $\in \{ 0,1\}$. Born process ($2\to 2$). 
\item {\tt ivirt}: integer $\in \{ 0,1\}$. Virtual correction ($2\to 2$). 
\item {\tt ifina}: integer $\in \{ 0,1\}$. Final state
  singularities ($2\to 2$).
\item {\tt iaini}: integer $\in \{ 0,1\}$. Initial state
  singularities on side $a$, i.e., the virtual photon side ($2\to
  2$). For calculating a DIS (direct) cross section, {\tt iaini}$=0$
  and {\tt iqcut}$=0$ (see below), since no initial state singularity
  for the virtual photon has to be subtracted. For the D and SR
  contributions, {\tt iaini} has to be calculated separately, due to
  different kinematics than the other $2\to 2$ processes. 
  In photoproduction and for the DR contributions, set {\tt iaini}$=1$
  with a finite value of {\tt cutmin} (see below). For calculating the
  virtual photon splitting term in DIS, calculate {\tt iaini}
  separately with {\tt cutmin}$=0.d0$ (see below). [sec 2.3]
\item {\tt ibini}: integer $\in \{ 0,1\}$. Initial state
  singularities on side $b$  ($2\to 2$), i.e., for the real photon or
  the hadron, according to {\tt iproc}.
\item {\tt iinco}: integer $\in \{ 0,1\}$. NLO $2\to 3$ contribution
  inside jet cone. Setting $R=0$ (see below) will suppress these
  contributions. 
\item {\tt iouco}: integer $\in \{ 0,1\}$. NLO $2\to 3$ contribution
  outside jet cone.
\end{itemize}
For a LO calculation, only {\tt iborn}$=1$. For a complete NLO
calculation, all parameters {\tt iborn, ivirt,$\ldots$,iouco} have to
be set equal to $1$ (modulo the changes for DIS or photoproduction
cross sections or for calculating the $\g^*\to q\bar{q}$ subtraction
term). 
\begin{itemize}
\item {\tt iqcut}: integer $\in \{ 0,1\}$. Insert a finite cut-off
  $y_s$ ({\tt cutmin}) into the integration on the virtual photon side
  for the $2\to 3$ NLO contributions ($=1$) or not ($=0$). See remarks
  for {\tt iaini}. [sec 2.3]
\end{itemize}

\subsection{Initial State}

\begin{itemize}
\item {\tt Ea, Eb}: real*8. Energies of the incoming particles $a$ and
  $b$ in GeV. Particle $a$ is always the lepton on the virtual photon
  side, whereas particle $b$ is a hadron for {\tt iproc}$=1$ and a
  lepton for {\tt iproc}$=2$.
\item {\tt iframe}: integer $\in \{ 0,1\}$. Defines the frame of
  reference, in which the cross sections are calculated. Setting 
  ${\tt iframe}=0$ selects the hadronic c.m.s. Setting ${\tt
  iframe}=1$ selects the laboratory frame (only correct for
  photoproduction, i.e., $Q^2\simlt 1$ GeV$^2$).
\item {\tt izaxis}: integer $\in \{ -1,1\}$. Defines the direction of
  the incoming electron. For {\tt izaxis}$=1$ the electron is traveling in
  the positive $z$-direction, for {\tt izaxis}$=-1$ it is traveling
  into the negative $z$-direction. 
\item {\tt iQ2}: integer $\in \{ 0,1,2,3,\ldots \}$. For {\tt
  iQ2}$=0$, $Q^2$ is integrated in $[{\tt Q2min},{\tt
  Q2max}]$. For {\tt iQ2}$=1$, $d\si/dQ^2$ is calculated at {\tt Q2min}. For
  {\tt iQ2}$>1$, $d\si/dQ^2$ is calculated at $n={\tt iQ2}$
  equidistant $Q^2$ points in $[{\tt Q2min},{\tt Q2max}]$. 
\item {\tt Q2min, Q2max}: real*8 $\ge 0$. Minimum and maximum value of
  $Q^2$. Setting ${\tt Q2min}=0$ will produce a photoproduction cross
  section. Remember setting {\tt iqcut}$=1$ for photoproduction cross
  sections.
\item {\tt iypol}: integer $\in \{ 1,2,3\}$. Polarization of the
  virtual photon. \\ $1=$ transverse (T), $2=$ longitudinal (L), $3=$T+L.
\end{itemize}

\subsection{Subprocess}

\begin{itemize}
\item {\tt W2min}: real*8. Minimum value of hadronic c.m.s.\ energy.
\item {\tt xhmin, xhmax}: real*8 $\in [0,1]$. Minimum and maximum
  value of $x_{Bj}$. 
\item {\tt Nf}: real*8 $\in \{ 1,2,3,4,5\}$. Number of active
  flavours. 
\item {\tt lambda}: real*8 $>0$ in GeV. Value of $\Lambda_{QCD}$ for
  {\tt Nf} flavours. 
\item {\tt ialphas}: integer $\in \{ 1,2\}$. One-loop ({\tt
    ialphas}$=1$) or two-loop ({\tt ialphas}$=2$) formula for the 
  strong coupling constant $\al_s$ without thresholds.  For ${\tt
    ialphas}=3$ the value of $\al_s$ is  taken from the {\tt PDFLIB}
  (automatically adjusts $\Lambda$).
\item {\tt icut}: integer $\in \{ 1,2,3,\ldots \}$. Phase space
  slicing parameter $y_s$. For {\tt icut}$=1$, $y_s$ is set equal to 
  {\tt cutmin}. For {\tt icut}$>1$, the cross section will be
  calculated at $n={\tt icut}$ points with logarithmic spacing in 
  the range $[{\tt cutmin},{\tt cutmax}]$.  
\item {\tt cutmin, cutmax}: real*8 $>0$. Minimum and maximum value
  of $y_s$. The independence of the NLO cross sections on $y_s$ has
  been tested for $y_s\in [10^{-2},10^{-4}]$. Large statistical errors
  occur for too small $y_s$. Recommended value: $y_s=10^{-3}$. 
\item {\tt fproc}: integer $\in \{ 1,2,3\}$. Selection of partons from
  PDF's. $1=$gluon, $2=$quarks, $3=q+g$.
\item {\tt idisga}: integer$\in [0,1]$. Since all partonic cross
  sections in \jv\ are implemented in the
  $\overline{\mbox{MS}}$-scheme, a photon PDF constructed in the
  DIS$_\g$ scheme has to be transformed into the
  $\overline{\mbox{MS}}$-scheme. This is done for the photon on side
  a by setting {\tt idisga}$=1$. [sec 2.4, page 14]
\item {\tt xaobsmn, xaobsmx}: real*8 $\in [0,1]$. Minimum and maximum
  value of $x_\g^{obs}$ for the virtual photon. 
\item {\tt ipdftyp}: integer $\in \{ 1,2,3,4\}$. Selects the PDF for the
  resolved virtual photon. \\ $1=$ {\tt SaS}, $2=$ {\tt GRS}, $3=$
  {\tt DG}, $4=$ {\tt PDFLIB} (for real photons with $Q^2=0$). 
\item {\tt igroupa}: integer. For {\tt ipdftyp}$=3,4$ this selects the
  group from the {\tt PDFLIB} (see manual \cite{pdflib}). For 
  {\tt ipdftyp}$=1$ this represents {\tt isasset}$\in \{
  1,2,3,4\}$, which selects input scale and scheme of SaS virtual
  photon PDF (see \cite{19}).
\item {\tt iseta}: integer. For {\tt ipdftyp}$=3,4$ this selects the
  set from the {\tt PDFLIB} (see manual \cite{pdflib}). For 
  {\tt ipdftyp}$=1$ this represents {\tt isasp2}$\in \{
  1,2,3,4,5,6,7\}$, which selects the scheme used to evaluate
  off-shell anomalous component in SaS virtual photon PDF (see \cite{19}).
\item {\tt idisgb}: integer $\in \{ 0,1\}$. See {\tt idisga}. 
\item {\tt xbobsmn, xbobsmx}: real*8 $\in [0,1]$. Minimum and maximum
  value of $x_\g^{obs}$ for real photon (only in
  $e\g$-scattering). [not yet implemented!]
\item {\tt igroupb}: integer. Selects the group from {\tt PDFLIB} for the
  resolved component for particle b (see manual \cite{pdflib}). {\tt iproc}
  automatically selects, if a hadron or a photon PDF is used. 
\item {\tt isetb}: integer. Selects the set from {\tt PDFLIB} (see
  manual \cite{pdflib}). 
\item {\tt a, b, c}: real*8 $>0$. Choosing the overall-scale $\mu^2$
  according to $\mu^2=a + bQ^2 + cp_T^2$. 
\end{itemize}

\subsection{Final State}

\begin{itemize}
\item {\tt jr}: real*8 $\ge 0$. Jetradius for cone algorithm. [sec 2.5]
  \\ By setting {\tt jr}$=0$, the function {\tt kincut} (see below)
  gives the kinematic variables of the partons, rather than those of
  the jets. Thus, other jet recombination schemes, apart from the cone
  algorithms, can be implemented. 
\item {\tt Rsep}: real*8 $>0$. The $R_{sep}$ parameter for the 
  modified Snowmass cone algorithm. Setting $R_{sep}=R$ (i.e. 
  {\tt jr}={\tt Rsep}) selects the $k_\perp$ algorithm. [sec 2.5]
\end{itemize}
The following variables are the $E_T$'s and $\eta$'s of the final state
jets. The final state phase space $d\mbox{PS}^{(n)}$ is parametrized
by these variables, so that a restriction on the $E_T$'s and $\eta$'s 
does not reduce the statistics. All {\tt VEGAS}-points are thrown
inside the given limits. If the limits are outside the physical
ranges, the program will automatically set the physically possible
limits for the variables, as they follow from energy-momentum
conservation. The terminology is the same as used for {\tt iQ2}, 
{\tt Q2min}, {\tt Q2max}:
\begin{itemize}
\item {\tt ipt}: integer $\in \{ 0,1,2,3,\ldots \}$. Transverse energy
  $E_{T_1}$ of the trigger jet. This is not necessarily the hardest
  jet. The transverse energy of the second jet $E_{T_2}$ is accessible
  in the user defined routine {\tt kincut} (see below). 
\item {\tt ptmin, ptmax}: real*8 $>0$. Minimum and maximum value of
  $E_{T_1}$.
\item {\tt ieta1}: integer $\in \{ 0,1,2,3,\ldots \}$. Rapidity
  $\eta_1$ of the trigger jet. 
\item {\tt etamin1, etamax1}: real*8 $>0$. Minimum and maximum value of
  $\eta_1$.
\item {\tt ieta2}: integer $\in \{ 0,1,2,3,\ldots \}$. Rapidity
  $\eta_2$ of the second jet. 
\item {\tt etamin2, etamax2}: real*8 $>0$. Minimum and maximum value of
  $\eta_2$.
\end{itemize}

\subsection{{\tt Vegas} and Output}

\begin{itemize}
\item {\tt ipoint}: integer $>0$. Defines the statistics of the
  integration. Typical value for LO: ${\tt ipoint}=3000$. Gives
  accuracy of under $1\%$. Typical value for NLO direct ${\tt
  ipoint}=100000$, for NLO resolved ${\tt ipoint}=150000$. Gives
  accuracy of around $3\%$ (the accuracy also depends on the selected
  value of the slicing parameter $y_s$. For smaller $y_s$, larger
  compensations occur and therefore the statistical errors become larger). 
\item {\tt itt}: integer $>0$. Number of iterations for {\tt Vegas}. 
  Recommended value: {\tt itt}$=5$.
\item {\tt iprn}: integer $\in \{ 0,10\}$. For ${\tt iprn}=0$, no
  output will be given. For ${\tt iprn}=10$, the result of the
  iterations of {\tt Vegas} will be printed on the screen. 
\item {\tt eps}: integer $>0$. Precision of the {\tt Vegas}
  integration. Recommended value: {\tt eps}=$10^{-5}$.
\item {\tt jfileout}: character*20. Name of the output-file. 
\end{itemize}

\subsection{The User-defined Function {\tt kincut}}

To allow a high flexibility for incorporating specific types of jet
selection processes and kinematic cuts, characterizing certain
experiments, the user can edit the function {\tt kincut.f}. The
following variables are {\tt common} and therefore accessible by the
user: 
\begin{center}
{\tt  y, x-bj, Q2, sH, elek, prot, Nc, Cf, pi, me } 
\end{center}
The jet-variables $E_T$, $\eta$ and $\phi$, are stored in the
three-dimensional vectors {\tt jet1(i)}, {\tt jet2(i)} and {\tt
  jet3(i)} for {\tt i=1,2,3}. The first entry 
of each vector gives the transverse energy, the second gives the
rapidity and the third gives the azimuthal angle. If there are only
two jets in the final state, the vector {\tt jet3} contains the value
$-10^{5}$ in each entry. Note, that the first and second jet are
not ordered in $E_T$. However, the transverse energy of the third jet
is always smaller than the transverse energy of the other two, 
$E_{T_3}<\mbox{min}(E_{T_1},E_{T_2})$. 

If the kinematic cuts are passed, {\tt kincut} returns {\tt 0},
otherwise {\tt 1}.

\section{Using \jv\ on Computers}

\subsection{Environment and Installation}

\jv\ is programmed in standard {\tt FORTRAN 77}. The
multidimensional phase-space integration is performed with help of the
{\tt Vegas} \cite{25} Monte-Carlo integration routine. Further
packages used are the {\tt PDFLIB}, {\tt SaSGAM} and {\tt GRS} PDF's. 

The \jv\ package is accessible via www.\footnote{The www-page
  is: http://www.desy.de/\~{}poetter/jetvip.html.} The source-code has
been translated and tested for the {\tt HP-UX}, {\tt IRIX (SGI)} and
{\tt Linux} computing platforms. Other platforms can be made available
upon request from the author. The package is tarred and can be
installed by doing {\tt tar xf name.tar}. It contains the following
files:
\begin{itemize}
\item Makefile: Typing {\tt make} will produce an executable JetVIP. Check
  the Makefile to make sure, the path for the PDFLIB's is correct (the
  standard path is {\tt /cern/pro/lib}).
\item common.f: Contains all common variables.
\item jv-master.o: The object-file of the main program \jv. 
\item kincut.f: A function, where the user can implement
     specific cuts on jet variables.
\item lib{\tt Vegas}-\${\tt ARCH.a}: The {\tt Vegas} library,
  containing the Monte-Carlo integration routine for the architecture
  {\tt ARCH}.
\item st.in: An example of the input-file. It will produce a LO calculation
of the $E_T$-spectrum of the direct reaction for $eP$-scattering at HERA.
The results are written to the output-file {\tt test.out}.
\end{itemize}
For running \jv\ with the example file, type {\tt JetVIP st.in [return]}.

\subsection{Typical Running Times}

The running times for LO and NLO calculations are very different. If we
start with a LO calculation, the benchmarks listed in Tab.~\ref{bench}
exist for different machines (as an example calculation we have chosen
the steering file shown in the appendix, which calculates the SR
component).  When the DR processes are considered, the running times
are typically 3--4 times longer. 

In a NLO calculation, the two-body and three-body processes are
calculated separately. The two-body processes are large and negative
and the three-body processes large and positive. Therefore, adding
both contributions to have the result will produce large compensations
between the two components. The statistical error from the {\tt Vegas}
integration, which is small for each component alone, is then large
compared to the sum of the two components. Therefore, the parameter
{\tt ipoint} in the steering file has to be a factor of 50 larger for
a NLO calculation, than for LO calculation, which makes the NLO
calculation rather slow. 

\begin{table}[ttt]
\renewcommand{\arraystretch}{1.2}
\caption{\label{bench}Benchmarks for a DIS calculation at HERA in LO
  with \jv.}
\begin{center}
\begin{tabular}{|l|l|r|} \hline
{\bf HP-UX} (translated on HP-UX9) & HP-UX9 & 19.1 s \\ 
                                   & HP-UX10 & 14.0--24.2 s \\ \hline
{\bf IRIX} (translated on IRIX5.3) & IRIX5.3 & 63.8 s \\ 
                                   & IRIX6.2 & 38.1 s \\ \hline
{\bf Linux} (translated on Pentium II) & 133MHz & 40.8 s \\ 
& 333 MHz & 11.6 s \\ \hline
\end{tabular}
\end{center}
\renewcommand{\arraystretch}{1}
\end{table}

Typically, calculating a NLO SR cross section with fixed bin-size
in one of the kinematical variables will take around 30 minutes. The
DR contributions will take from 2 to 4 hours, due to the large number
of matrix elements that have to be considered in the calculation.

\subsection*{Acknowledgments}

I have profited very much from discussions with D.~Graudenz,
M.~Klasen, T.~Kleinwort and G.~Kramer. I thank T.~Kleinwort for the
consent in using his implementation of the resolved-resolved matrix
elements and for his support concerning technical questions of the
computing environment. Several comments from users have improved the
code of the program. In particular I am grateful to S.~Mattingley,
T.~McMahon, M.~Tasevsky, J.H.~Vossebeld and M.~Wobisch. I thank
G.~Kramer for comments on the manuskript. 

%% file: liter.tex

%% file: appendix.tex
\section*{Appendix A: Input-file}

The following example of an input file will produce an $E_T$ spectrum
of the direct component of the single-jet cross section $d\si /dE_T$
in $eP$-collisions under typical HERA conditions in LO in the hadronic
c.m.s., using the scale $\mu^2=Q^2+E_T^2$. 

\begin{small}
\begin{tt}
\begin{tabbing}
'========================================================================='\\
'       \qquad\qquad \=       $\!\!\!\!\!\!\!\!\!$  CONTRIBUTION' \=  \\
'========================================================================='\\
1       \>      iproc   \>[1=ep; 2=ee]\\
1       \>      ijet    \>[1=Single Jet; 2=Dijet]\\
2       \>      isdr    \>[1=D; 2=SR; 3=SR*; 4=DR]\\
1       \>      iborn   \>[born]\\
0       \>      ivirt   \>[virtual]\\
0       \>      ifina   \>[final state sing.]\\
0       \>      iaini   \>[initial state sing. side a]\\
0       \>      ibini   \>[initial state sing. side b]\\
0       \>      iinco   \>[2-->3 inside cone]\\
0       \>      iouco   \>[2-->3 outside cone]\\
0       \>      iqcut   \>[yqi-min in 2->3 matrices? 1=yes; 0=no]\\
'========================================================================='\\
'       \>      INITIAL STATE'\> \\
'========================================================================='\\
27.5d0  \>      Ea      \>[energy on side a (ep,ee: lepton)]\\
820.d0  \>      Eb      \>[energy on side b (ep: proton; ee: lepton)]\\
'------- Frame of reference ----------------------------------------------'\\
0       \>      iframe  \>[frame of ref: 0=hadr CMS; 1=HERA-Lab]\\
1       \>      izaxis  \>[lepton "a" travels in pos (=1) or neg (=-1) z-dir]\\
'------- Lepton a --------------------------------------------------------'\\
0       \>      iQ2     \>[0=si; 1=dsi/dQ2; >1=dsi/dQ2(Q2)]\\
5.d0    \>      Q2min   \>[0.d0 selects photoproduction]\\
11.d0   \>      Q2max   \>\\
0       \>      iwwa    \>[which Weizs.-Will.: 0=ln(Q2mx/Q2mn); 1=ln(thmax/me)]\\
0.d0    \>      thetamn \>[min angle of deflection for lepton in DIS]\\
180.d0  \>      thetamx \>[max angle]\\
11.d0   \>      Ea-min' \>[min energy of outgoing lepton]\\
0.05d0  \>      ymin    \>[min y]\\
0.6d0   \>      ymax    \>[max y]\\
3       \>      iypol   \>[Photon polarization: 1=T, 2=L, 3=L+T]\\
'------- Lepton b (only relevant for the ee-case ) -----------------------'\\
4.d0    \>      P2max   \>[P2=virtuality of real photon]\\
0       \>      iwwb    \>[which Weizs.-Will.: 0=ln(P2mx/P2mn); 1=ln(thmax/me)]\\
180.d0  \>      thetbmx \>[max angle for Weizs.-Will]\\
\end{tabbing}
\end{tt}
\end{small}
\newpage

\begin{small}
\begin{tt}
\begin{tabbing}
'========================================================================='\\
'       \qquad\qquad \=       $\!\!\!\!\!\!\!\!\!$  SUBPROCESS' \=  \\
'========================================================================='\\
0.d0    \>      W2min   \>[min hadronic cms energy]\\
0.d0    \>      xhmin   \>[min x-bjorken]\\
1.d0    \>      xhmax   \>[max x-bjorken]\\
5.d0    \>      Nf      \>[Number of active flavours]\\
0.204d0 \>      lambda  \>[Lambda-QCD (has to match Nf)]\\
2       \>      ialphas \>[QCD coupling: 1=one-loop; 2=two-loop; 3=PDFLIB]\\
1       \>      icut    \>[Slicing param. y-cut: 1=fixed; >1=si(y-cut)]\\
1.d-3   \>      cutmin  \>      \\
1.d-3   \>      cutmax  \>      \\
'------ PDFs for the resolved contributions ------------------------------'\\
3       \>      fproc   \>[Parton from PDF: 1=g, 2=q, 3=q+g]\\
0       \>      idisga  \>[DISg -> MSbar for photon a]\\
0.d0    \>      xaobsmn \>[photon a - minimal xa-gamma obs]           \\
1.d0    \>      xaobsmx \>[max xa-gamma obs]                          \\
1       \>      ipdftyp \>[PDF for y*(res): 1=SaS; 2=GRS; 3=DG; 4=PDFLIB]\\
1       \>      igroupa \>(Param. for SaS or PDFLIB --> see manual)   \\
2       \>      iseta   \>( " )                                       \\
1       \>      idgsas  \>( " )                                       \\
0.d0    \>      xbobsmn \>[photon b - minimal xb-gamma obs]\\
1.d0    \>      xbobsmx \>[max xb-gamma obs]\\
4       \>      igroupb \>[authors of PDF on side b: y(res) or prot]\\
34      \>      isetb   \>[Set-No]\\
'------- Scales ----------------------------------------------------------'\\
0.d0    \>        a     \>  [Scale: mu**2=a+b*Q**2+c*pt**2]\\
1.d0    \>        b     \>  \\
1.d0    \>        c     \>  \\
'========================================================================='\\
'       \>        FINAL STATE'\>\\
'========================================================================='\\
1.d0    \>        jr   \>   [Jetradius for Snowmass convention] \\
2.d0     \>       Rsep \>   [R-sep parameter]\\
'------ Jet variables ----------------------------------------------------'\\
11      \>        ipt    \> [0=si; 1=dsi/dpt; >1=dsi/dpt(pt)]\\
5.d0    \>        ptmin \>  [pt of trigger jet]\\
15.d0   \>        ptmax  \>         \\
0       \>        ieta1  \> [0=si; 1=dsi/deta1; >1=dsi/deta1(eta1)]\\
-30.d0  \>        eta1min \>[eta of trigger jet]\\
30.d0   \>        eta1max\>\\
0       \>        ieta2  \> [0=si; 1=dsi/deta2; >1=dsi/deta2(eta2)]\\
-30.d0  \>        eta2min \>[eta of second hardest jet]\\
30.d0   \>        eta2max\>\\
\end{tabbing}
\end{tt}
\end{small}

\newpage

\begin{small}
\begin{tt}
\begin{tabbing}
'========================================================================='\\
'       \qquad\qquad \=       $\!\!\!\!\!\!\!\!\!$   VEGAS and OUTPUT' \=  \\
'========================================================================='\\
'------- VEGAS parameters ------------------------------------------------'\\
3000     \>       ipoin  \> [No. of points per integr.]\\
5        \>       itt    \> [No. of iterations]\\
10       \>       iprn   \> [Printflag: 10=screen output]\\
1.d-5    \>       eps    \> [Precision of int.]\\
'------- Name of output file ---------------------------------------------'\\
'test.out' \>     jfileout \>[Filename]\\
\end{tabbing}
\end{tt}
\end{small}

\section*{Appendix B: Output-file}

If the user chooses to calculate the spectrum of a variable, the
output-file will contain the result in the following form:
\begin{small}
\begin{tt}
\begin{tabbing}
 5.0 \qquad\qquad \= 1026.109741210937 \qquad \=  4.70078134536743\\
 6.0 \>513.3152465820313 \>2.41036343574524\\
 7.0 \>284.2205505371094 \>1.30083477497101\\
 8.0 \>164.3175811767578 \>.7139108777046204\\
 9.0 \>101.2273483276367 \>.4778541326522827\\
 10.0 \>64.58644104003906 \>.2957088351249695\\
 11.0 \>42.4879379272461 \>.1964362561702728\\
 12.0 \>29.31233406066895 \>.1275904476642609\\
 13.0 \>20.39706230163574 \>9.869953989982605E-02\\
 14.0 \>14.49518013000488 \>6.361690163612366E-02\\
 15.0 \>10.62311935424805 \>4.653203114867210E-02
\end{tabbing}
\end{tt}
\end{small}
The first entry is the point in the spectrum (in this case the value
of $E_T$), the second entry is the resulting cross section in [pb]
(here $d\si /dE_T$) and the third entry gives the statistical errors
of the {\tt VEGAS}-integration in [pb].

If the user chooses to calculate a certain bin (by setting all
integration variables, i.e., {\tt iQ2}, {\tt ipt}, {\tt ieta1} and 
{\tt ieta2}, equal to zero), the output-file will contain only one
line, with the first number being zero:
\begin{small}
\begin{tt}
\begin{tabbing}
 .0 \qquad\qquad \= 1691.092895507812 \qquad \= 9.78220748901367 \\
\end{tabbing}
\end{tt}
\end{small}
The second entry again gives the cross section in the chosen bin in
[pb] and the third entry gives the error. Note that the cross section is
not normalized by the bin width.